# World-line observables and clocks in General Relativity

Hossein Farajollahi

School of Mathematics and Statistics

University of Sydney, NSW 2006, Australia

hosseinf@maths.usyd.edu.au

**Abstract**

A proposal for the issue of time and observables in any parameterized theory such as general relativity is addressed. Introduction of a gauge potential 3-form A in the theory of relativity enables us to define a gauge-invariant quantity which can be used by observers as a clock to measure the passage of time. This dynamical variable increases monotonically and continuously along a world line. Then we define world line observables to be any covariantly defined quantity obtained from the field configurations on any such causal past with dynamical time T.

# Introduction

The problem of evolving of a dynamical system from initial data is known as the Cauchy problem or initial value problem **(1)**. In General Relativity, the Cauchy problem is naturally addressed using the 3+1 ADM representation.

In the ADM approach, the spatial hypersurface $\Sigma$ is assumed to be equipped with a space-like 3-metric $h_{ij}$ induced from space-time metric $g_{\mu\vartheta}$.

Einstein's equations $G_{\mu\vartheta} = R_{\mu\vartheta} - \frac{1}{2} R g_{\mu\vartheta} = 0$, are of course covariant and do not single out a preferred time with which to parametrise the evolution. Nevertheless, we can specify initial data on a chosen spatial hypersurface $\Sigma$, and if $\Sigma$ is Cauchy, we can evolve uniquely from it to a hypersurface in the future or past. The issue of specification of initial or final data on Cauchy hypersurfaces has been discussed in many papers; for example, see **(2)**.

We regard this 3-metric as the fundamental variable, and specify $h^{ij}$ and momenta $\pi^{ij}$ on $\Sigma$. Instead of the momenta $\pi^{ij}$, we could if we wished specify the extrinsic curvature $K^{ij}$ given by $K^{ij} = -|h|^{-1/2} (\pi^{ij} - \frac{1}{2} h^{ij} \pi_l^{\ l})$

However, one may fix the initial data on null hypersurfaces rather than spatial hypersurfaces. This is known as the characteristic initial value problem.

In General Relativity it is natural to work with a foliation of space-time by space-like hypersurfaces, as this reflects the older Newtonian idea of a 3-dimensional universe developing with time. This seems close to our experiences and is easy to visualize. Nevertheless, null hypersurfaces and null directions should be considered for the following reasons:

- The procedure of determining 'initial conditions' on spacelike hypersurface is unrealistic and unnatural in the context of relativity **(3)**. This is because no information can be obtained from space-time points which are separated by space-like distances. In particular, an observer has access only to information originated from his past light cone **(4)**. This is an immediate consequence of the laws of relativity, if we assume that physical observations are made by a single localized observer.

- There has been considerable success in using null boundaries to formulate the canonical theory of gravitational radiation on outgoing null surfaces. This is because in electromagnetism and gravitation (which are mediated by particles with zero mass), fields propagate in null directions and along null hypersurfaces **(5)**.

- In some cosmological models of interest, space-time is not globally hyperbolic and so there are no Cauchy hypersurfaces on which to specify boundary data. In such cases, data specified on a space-like hypersurface cannot be used to generate a unique classical solution and therefore cannot be used to label a particular point in the phase space. Even if the space-time is globally hyperbolic, it may not be possible for localized observers to gather all the necessary boundary data from a space-like Cauchy hypersurface. Indeed, unless the space-time is deterministic, there will be no event whose casual past contains the hypersurface. In this case, no localized observer will have access to enough data to distinguish between different classical solutions - i.e. between different elements of the phase space.

- The formulation of gravitational radiation field on the null surface lays bare the dynamical degrees of freedom in the theory and allows one to analyze the properties of the gravitational radiation field in terms of these quantities **(6)**.

The approach of setting the final data on a null hypersurface is essential if we are interested in a theory of observations made by a single localized observer who can collect observational data only from that subset of space-time which lies in his causal past.

The set of observables measured by any observer contain a lot of redundant information. A successful theory is one that accurately describes the relationships between these observables so that information about the universe can be deduced from as small set of data as possible. Thus, it is natural to look for a minimal set of observables originating from the observer's past light cone which can be used to reconstruct the values of all other observables. The minimal set of observables on the past light cone is required to be complete in the sense that they can be used to determine the values of all the other observables in the causal past **(7)**.

The aim of this paper is to identify a set of observables in General Relativity on the past light cone of a single localized observer. We also introduce a volume clock which can be used to measure the evolution of these observables.

## 1 Dirac observables in General Relativity

General Relativity, like many other field theories, is invariant with respect to a group of local symmetry transformations **(8)**. The local symmetry group in General Relativity is the group Diff (*M*) of diffeomorphisms of the space-time manifold $M^1$.

---

[1] Here we consider only globally hyperbolic and orientable space-times.

In General Relativity, Dirac observables **(9)** must be invariant under the group of local symmetry transformations. The Hamiltonian constraint and momentum constraint in General Relativity are generators of the symmetry transformations, and so a function $\Phi$ on the phase space is a Dirac observable, iff

$$\{\Phi, H(x)\} = \{\Phi, H_i(x)\} = 0 \tag{1}$$

at all points $x \in M$. Such observables are necessarily constants of motion and are called *perennials*[2] by Kuchar. They are invariant under local Lorentz rotations $SO(3)$ and $Diff \, \Sigma$ (as well as $SO(1, 3)$).

However, according to Kuchar, there are serious problems in identifying and defining perennials **(10)-(11)**. Even if they exist, they are likely to be complicated functions on the constraint surface, difficult to define explicitly and to work with. In particular, it has been argued that perennials can not be identified in classical or quantum gravity so one should concentrate on formulating nonexistence theorems about them **(11)**.

---

[2] The name *perennials* are given by Kuchar to distinguish them from "observables" which are not necessarily constant of motion along classical trajectories. For instance, according to his definition, a dynamical variable $\Phi$ constructed from the metric field is a Kuchar observable only if its value is unaffected by spatial diffomorphisms,

$$\{\Phi, H_i(x)\} = 0.$$

Thus for example the volume of the spatial hypersurface $\Sigma$

$$V[g] = \int_\Sigma \sqrt{{}^3 g(x)} d^3 x,$$

is a Kuchar observable but not necessarily a constant of motion, since it does not necessarily commute with Hamiltonian constraint $H(x)$ **(11)**.

The above criteria for observables in relativity appear to rule out the existence of local observables if locations are specified in terms of a particular coordinate system. Indeed, it might appear that one would be left with only observables of the form

$$\Phi = \int \varphi(x)\sqrt{-g(x)}d^4x, \qquad (2)$$

where $\varphi(x)$ is an invariant scalar as for example $R$, $R^2$, $R^{\mu\vartheta}R_{\mu\vartheta}$. While such observables clearly have vanishing Poisson brackets with all the constraints, they can not be evaluated without full knowledge of the future and past of the universe. While this may be deducible in principle from physical measurements made at a specific time, it is well beyond the scope of any real experimenter. These observables are perennials, as they are necessarily constant of motion along classical solutions **(10)**.

In reality, observations are made locally. We therefore ought to be able to find a satisfactory way to accommodate local observables within General Relativity. In particular, we would like to be able to talk about observables measured at a particular time, so that we can discuss their evolution. Local observables in classical or quantum gravity must be invariant under coordinate transformations. The difficulty in defining local observables in classical gravity is that diffomorphism invariance makes it difficult to identify individual points of the space-time manifold **(12)**.

It is fairly easy to construct observables which commute with the momentum constraints. Such observables can be expressed as functions of dynamical variables on the spatial hypersurfaces. However, according to the Dirac prescription, observables must also commute with Hamiltonian constraint. This requirement is much harder to satisfy, and leaves us with just Kuchar's perennials.

However, in a slightly different formalism, Rovelli addressed the problem by introducing a Material Reference System (*MRS*) **(13)**. By *MRS*, Rovelli means an ensemble of physical bodies, dynamically coupled to General Relativity that can be used to identify the space-time points.

In this approach, all the frames and all the test particles are assumed to be material objects. However, to implement the process and simplify the calculation, one has to neglect the energy-momentum tensor of matter fields in the Einstein equations, as well as their contributions to the dynamical equations for matter fields **(14)**. Of course the price that one has to pay for this neglect is obtaining an indeterministic interpretation of the Einstein equations. General Relativity is then approximate[3] because we disregard the energy-momentum of the MRS as well as incomplete (because we disregard dynamical equations of the MRS **(15)**. However, the indeterminism here is not fundamental and does not imply that Dirac determinism is violated **(13)**. In fact this approximation can arise in any field theory and has always been resolved by considering a limiting procedure in which the rest masses, charges, etc., of test bodies tends to zero **(16)**.

Rovelli's observables can be interpreted as the values of a quantity at the point where the particle is and at the moment in which the clock displays the value *t*. However *t* itself is not an observable, even though its conjugate momentum is constant along each classical trajectory.

By introducing a cloud of particles filling space, with a clock attached to every particle, one can easily generalize the model to a continuum of reference system particles, in order to get a complete material coordinate system and a complete set of physical observables.

---

[3] In quantum theory, the operators correspond to these observables obey an approximate Heisenberg equation **(17)**.

Rovelli's 'evolving constants of motion' are genuine Dirac's observables. They are constant of motion since they commute with Hamiltonian and momentum constraints, while evolving with respect to the clock time $t$.

Rovelli's observables are functions defined on spatial hypersurfaces. He assumes the space-time has a topology $\Sigma \times R$ where $\Sigma$ is a compact spatial hypersurface and $R$ is the real time. In order to have evolution into the future or past the spatial hypersurface must be a Cauchy hypersurface. This makes sense if the underlying space-time is assumed to be globally hyperbolic.

Perhaps more importantly, the observations collected by the observers will not generally be accessible to any single observer, and so Rovelli's approach is not useful if we set a theory of observations by a single observer.

However, in some cosmological models of interest, space-time is not globally hyperbolic and so there are no Cauchy hypersurfaces. In such cases, one cannot identify observables on spatial hypersurfaces.

## 2  The causal structure of General Relativity

Consider the space-time $M$ which is a 4-dimensional manifold with a smooth

Metric $g_{\mu\vartheta}$ having signature (-, +, +, +). Assume that $M$ is time oriented, and there is a global time-like vector field on $M$.

Giving any set $S \subset M$, the chronological future $I^+(S)$ of $S$ is defined as the set of all points in $M$ which can be reached from $S$ by a future-directed time-like curve in $M$ .The chronological past is defined by replacing 'future' by 'past' and the '+' by a '-'.

The causal future of $S$, $J^+(S)$, is defined as the union of $S$ with the set of all points which can be reached from $S$ by a future-directed non-space like curve in $M$. It is the region of space-time which can be causally affected by events in $S$. Similarly, we define $J^-(S)$, by replacing 'future' by 'past' and the '+' by a '-' **(2)**.

The future domain of dependence, or future Cauchy development of a closed set $S$, is the set $D^+(S)$ of all points $q \in M$, such that every past inextensible time-like curve through $q$ intersects $S$. A knowledge of the appropriate data on the set $S$ would determine events in a region $D^+(S)$ to the future of $S$. We define the past Cauchy development $D^-(S)$, similarly, the domain of dependence of $S$ is the union $D(S) = D^+(S) \bigcap D^-(S)$. In general, we suppose $S$ is some space-like or null 3-surface. The initial data on $S$ determines the entire evolution of $M$, past and future.

Closely related is the property of global hyperbolicity. If $M$ is globally hyperbolic, then for any two points $p, q \in M$ the set $C(q,p) = J^-(q) \bigcap J^+(p)$ is compact **(18)**. The property of global hyperbolicity is useful when considering hyperbolic differential equations on a manifold, and is the natural condition to ensure the existence and uniqueness of solutions **(19)**.

A globally hyperbolic space-time admits a Cauchy surface **(2)**; that is, a space-like hypersurface $\Sigma \subset M$ which is intersected by any inextensible causal curve in $M$ exactly once. (Thus, $M$ is in the domain of dependence $D(\Sigma)$ of the Cauchy hypersurface, $\Sigma$).

Any globally hyperbolic space-time $M$ has the simple topology $\Sigma \times R$ where $R$ is a real line and $\Sigma$ is some 3-manifold (Cauchy surface). Many physically interesting space-times fall into this category, such as flat Minkowski space-time, Schwarzschild space-time, and Robertson-Walker space-time **(19)**. A globally hyperbolic space-time also has a global time function, i.e. a function $T$ on $M$ such that $\nabla T$ is time-like, with the level sets of $T$ being Cauchy surfaces **(20)**.

Another closely related notion is determinism, which simply means that each event can predict its own future from its own past. The space-time $(M,g)$ is deterministic iff any past inextensible causal curve which intersects $I^+(q)$ must also intersect $I^+(q)$ **(21)**.

## 3  A dynamical clock for General Relativity

We suppose space-time is equipped with a smooth Lorentzian metric $g$ and a smooth gauge potential 3-form $A$ with field strength $F = dA$ satisfying the field equation

$$d^*F = 0. \qquad (3)$$

Note that this field strength $F$ is invariant under the gauge transformation $A \mapsto A + dB$ for any smooth 2-form $B$.

The field equation implies that the pseudo-scalar $\phi = {}^*F$ is constant throughout space-time. The square of this pseudoscalar behaves exactly as a cosmological constant if is coupled to the metric **(22)**.

The presence of the field $A$ allows us to define a gauge-invariant dynamical time variable, which increases continuously along any future-directed non-spacelike curve. Given any non-empty achronal set $S$ and any $q \in I^+(S)$, we define

$$T(S,q) \equiv -\oint_{\partial C(S,q)} \frac{1}{\phi} A \qquad (4)$$

where $\partial C(S,q)$ is the boundary of $C(S,q)$, the intersection of the causal past of $q$ and the causal future of $S$; i.e.

$$C(S,q) = J^-(q) \cap J^+(S). \qquad (5)$$

Taking the constant factor $\frac{1}{\phi}$ outside the integral and applying Stokes Theorem, equation (4) gives

$$T(S,q) = \frac{1}{\phi} \int_{C(S,q)} dA$$
$$= \frac{1}{\phi} \int_{C(S,q)} {}^*\phi \qquad (since \quad dA = F = -{}^*\phi) \tag{6}$$

which is precisely the invariant 4-volume of $C(S,q)$.

Assume that there is Cauchy hypersurface $\Sigma$ in the causal past of the non-empty achronal set $S$. Then $q$ is in the domain of dependence of $\Sigma$.

By using Proposition 6.6.6 in **(2)**, one sees that $I^-(q) \cap J^+(\Sigma)$ is compact.

Since $J^+(S) \cap J^+(\Sigma)$, then $C(S,q)$ must be also compact. It follows that the boundary $\partial C(S,q)$ is also compact and hence the integral (4) exists.

Given some non-empty achronal set $S$ in the causal future of some Cauchy hypersurface $\Sigma$, we define a time function $\tau: I^+(S) \to R$ with $\tau(q) = T(S,q)$.

**Theorem**: *Suppose $(M,g)$ is a globally hyperbolic space-time. Then $\tau$ is a time function on $I^+(S)$; i.e. $\tau$ is continuous and strictly increasing along future directed causal curves in $I^+(S)$.*

**Proof**: It is obvious that $\tau$ is strictly increasing since the manifold is orientable and the metric is non-degenerate, and $\tau(q)$ just measures the volume of the past of the point $q$. The continuity of $\tau$ can be demonstrated by adapting an argument of Geroch **(19)**. Let $q$ be any point in $I^+(S)$, the chronological future of $S$; $q \in I^+(S)$. Then $\tau(q) = T(S,q)$ is finite and is strictly positive. If $\omega$ denotes the volume 4-form obtained from the metric (= $\sqrt{-g}d^4x$ in any right-handed coordinate system), we define a new volume element

$$dV = [\tau(q)]^{-1}\omega \qquad (7)$$

on the set $I^-(q) \cap J^+(S)$, which can now be viewed as a globally hyperbolic space-time with total volume equal to unity (as one sees by integrating the new volume element). Defining

$$V^-(r) = \frac{\tau(q) - \tau(r)}{\tau(q) + \tau(r)} \quad \text{for each } r \in I^-(q), \qquad (8)$$

one then can adapt Geroch's argument to show that since the interior of $M$ is globally hyperbolic, then $V^-$ is continuous everywhere, and throughout $I^-(q) \cap J^+(S)$ in particular. (For some of the properties of the volume functions $V^-(r)$, see **(19)**.) It follows that $\tau$ is also continuous throughout $I^-(q) \cap J^+(S)$. □

Since $\tau$ is continuous and strictly increasing along future-directed non-spacelike curves, it generates a natural foliation of space-time as a sequence of constant $\tau$ surfaces.

For our purposes, the essential point is that $\tau$ is a gauge-invariant dynamical quantity which can be used by observers as a clock to measure the passage of time. The time $\tau$ also has an interesting interpretation in unimodular gravity.

Admittedly it is not obvious how one might measure a 3-form on a 3-manifold. Also, there is quite a complicated relationship between $\tau$ and proper time. In a flat space-time, the time measured by this clock along a time-like geodesic is proportional to the 4*th* power of proper time. In general however, the variable measured by this clock may not be expressible as a simple function of proper time.

# 4 World-line observables

We suppose that the space-time contains a future-directed time-like geodesic $\Gamma$ representing the world-line of an observer, as well as the 3-form gauge potential $A$ discussed earlier.

We also suppose that the metric $g$ satisfies Einstein's equations which are assumed to include a contribution from the energy momentum tensor of the gauge field.

Such a space-time will be referred to as a "space-time-with-observer". Space times-with-observer which are diffeomorphic to each other are physically indistinguishable, and for our purposes are taken to be equivalent. Each equivalence class represents a solution of the classical equations of motion, and hence an element in the phase space of our theory.

As shown in the last section, the dynamical variable $\tau$ increases continuously along the world line $\Gamma$, starting from 0 at the achronal set $S$ and possibly approaching a maximum value $\tau_{max}$. (If $\tau$ increases without limit along $\Gamma$, we take $\tau_{max} = \infty$.) Given any particular value of $\tau$ in the range $(\tau, \tau_{max})$, there is a unique point $p_\tau$ on the curve $\Gamma$ at which this value is attained. For this particular value of $\tau$, we define $\partial C_\tau$ as the boundary of causal past of the point. The time parameter $\tau$ along $\Gamma$ can be extended to give a time-like coordinate on $M$. However, notice that the other three coordinates are still quite arbitrary.

A quantity derived from the values of the fields and their derivatives on $\partial C_\tau$, which are invariant under 3-dimensional coordinate transformations within $\partial C_\tau$, will be called a $\Gamma$-observable. These observables are any covariantly defined quantity obtained from the fields on $\partial C_\tau$ with $\tau \in (0, \tau_{max})$.

We may construct $\Gamma$-observables of the form

$$\Psi^\Gamma = \int_{\partial C_\tau} \psi \qquad (9)$$

where 3-form $\psi$ is any covariantly defined field or its derivative on $\partial C_\tau$.

$\Gamma$-observables do not have vanishing Poisson brackets with the local Hamiltonian constraints. They are not therefore Dirac observables, reflecting the fact that the prespecified foliation is not invariant under local time evolution. Similarly the dynamical time $\tau$ is not a Dirac observable.

Following Rovelli **(13)**, we can construct a set of 'evolving constants of motion' such as $\Psi_\tau^\Gamma(q)$ by using the dynamical time variable introduced in the last section to set the conditions on the $\Gamma$-observable. This observable can be interpreted as the value of a quantity on $\partial C_\tau$ at the instant the observer's clock shows the value $\tau$. These quantities commute with the integrated Hamiltonian constraints. However, we do not require them to commute with the Hamiltonian constraint at each point on $\partial C_\tau$.

## 5 Summary

From a realistic point of view and for the purpose of experimental work it has been recognized that observation is a local interaction between the observer and observable and that observable correspond to the physical data that can be collected from observer's past light cone.

A gauge-invariant dynamical time defined that can be used by observers as a clock to measure the passage of time. Rovelli's evolving constants of motion' constructed on the observer's past light cone are genuine Dirac observables on null hypersurfaces.

## 6 Acknowledgement

I would like to thank Dr hugh luckock for his help in the achievement of this work.